\documentclass[12pt,reqno]{amsart}
\usepackage{amsmath}
\usepackage{amssymb}
\usepackage{amsxtra}
\usepackage{amsthm}
\usepackage{color}
\usepackage{slashed}
\usepackage{tikz} 
\usepackage{color}

\usepackage{hyperref}
\hypersetup{
    colorlinks = true,
    colorlinks = false,
    linkcolor = blue,
    anchorcolor = red,
    citecolor = blue,
    filecolor = red,
    pagecolor = red,
    urlcolor = blue
}
\usepackage{graphicx}

\usepackage{latexsym}
\usepackage{graphicx}

\def\ee{\end{equation}}
\def\bea{\begin{eqnarray}}
\def\eea{\end{eqnarray}}

\def\bs{\bar s}
\def\p{\partial}
\def\bp{\bar\partial}

\newcommand{\szego}{Szeg\"o\ }

\renewcommand{\Re}{{\operatorname{Re}\,}}
\renewcommand{\Im}{{\operatorname{Im}\,}}

\renewcommand{\epsilon}{\varepsilon}

\newcommand{\kahler}{K\"ahler }

\newcommand{\PP}{{\mathbb P}}

\newcommand{\R}{{\mathbb R}}
\newcommand{\C}{{\mathbb C}}

\newcommand{\kcalomega}{\mathcal{K}_{[\omega_0]}}

\newcommand{\CP}{\C\PP}

\newcommand{\bcal}{\mathcal{B}}

\newcommand{\dcal}{\mathcal{D}}

\newcommand{\kcal}{\mathcal{K}}

\newcommand{\mcal}{\mathcal{M}}

\newcommand{\qcal}{\mathcal{Q}}

\newcommand{\tr}{{\rm tr}\,}

\title{Stability and integration over Bergman metrics}

\begin{document}

\author{Semyon Klevtsov$^1$ and Steve Zelditch$^2$}

\maketitle
{\small
\address{\it$^1$Mathematisches Institut, Universit\"at zu K\"oln, Weyertal 86-90, 50931 K\"oln, Germany}

\address{\it $^2$Department of Mathematics, Northwestern  University, Evanston, IL 60208, USA}
\vspace{.2cm}
}

\begin{center}

\email{\tt\footnotesize  klevtsov@math.uni-koeln.de, zelditch@math.northwestern.edu}
\end{center}

\begin{abstract}

We study partition functions of random Bergman metrics, with the actions defined by a  class of geometric functionals known as `stability functions'.    
We introduce a new stability invariant - the critical value of the coupling constant-- defined as  the minimal coupling constant for which the partition function converges.  It measures the minimal degree of stability of geodesic rays in the space  the Bergman metrics, with respect to the action. We calculate this critical value when the action is the $\nu$-balancing energy, and show that
$\gamma_k^{\rm crit} = k - h$ on a Riemann surface of genus $h$.

\end{abstract}

\tableofcontents

\section{Introduction}

The problem of defining  path integrals over all metrics on a manifold dates back to the seminal work of Polyakov \cite{P}, where the path integral over all two-dimensional Riemannian metrics was proposed. The Polyakov's Liouville theory is defined with the natural diff-invariant path integral measure deriving from the DeWitt-Ebin metric on the space of all metrics. This infinite-dimensional space is not flat (in fact it is positively curved), which leads to intractable regularization problems of the measure. For the Liouville theory this problem was circumvented in the Minkowski case by the light-cone quantization \cite{KPZ}, and in the Euclidean case by "mapping" the model into the framework of conformal field theory \cite{D,DK}. Nevertheless, it would be very interesting to define the path integral over the infinite dimensional space of metrics on a manifold specifically in terms of the geometry of this space. 

Independently of path integrals over metrics, a  mathematical framework has been developed  by  Yau-Tian-Donaldson 
to study existence of extremal metrics in K\"ahler geometry, see e.g. \cite{PS} for review. In \cite{FKZ2,FKZ4}, the authors have been employing  this framework in order to define path integrals over spaces $\kcalomega$ of \kahler metrics $\omega_{\phi}$  in a fixed class on a \kahler manifold $(M, \omega_0)$ of complex dimension $n$. Mabuchi
(and later Semmes and Donaldson \cite{Don1}) have defined a Riemannian metric on $\kcalomega$ which at least formally makes it an
infinite dimensional  non-positively
curved symmetric space. In  a sense it is parallel to the positively curved deWitt-Ebin metric (known in the K\"ahler setting as the Calabi metric), and 
one would expect a genuine integral over $\kcalomega$ with respect to the Mabuchi volume form to be intractable as well. The main idea of our work is to define the integral as a limit of  finite dimensional integrals over spaces of so-called Bergman metrics $\bcal_k$. These metrics are induced by embedding $M$ into complex projective space $\CP^{N_k-1}$ using the space $H^0(M, L^k)$ of holomorphic sections of an ample line bundle $L^k\to M$  and pulling back to $M$ the Fubini-Study metric on $\CP^{N_k-1}$. Here, $N_k= \dim H^0(M, L^k)$ is the dimension of the space of holomorphic sections. 
The Bergman metric spaces $\bcal_k \subset \kcalomega$ are known to approximate $\kcalomega$ in a very strong asymptotic
sense. Moreover, if one chooses a reference metric $\omega_0$, then $\bcal_k$ is naturally identified with the  symmetric space $SL(N_k, \C)/SU(N_k)$ of positive definite Hermitian matrices of rank $N_k$. The partition function then has the general form, 
\begin{equation}\label{MPI}
Z_k(\gamma_k) = \int_{\bcal_k} e^{- \gamma_kS_k(\omega_0, \phi)} \dcal_k \phi,
\end{equation}
where $S_k(\omega_0, \phi): \bcal_k\to\mathbb R$ is an action functional, $\dcal_k \phi$ is an appropriate measure on $\bcal_k$ and $\gamma_k$ is the coupling constant.  There is a natural $SL(N_k, \C)$ action on $\bcal_k$ and in this
article we choose $\dcal_k \phi$ to be the invariant Haar measure. (The invariant metric and Haar measure are only
invariant up to scale and we will also consider rescaled measures $\dcal^{\epsilon_k} \phi_k$).  We also choose $S_k$ to be a geometric  `stability function' in the sense of  GIT (geometric invariant theory).  We focus in particular on a simple functional known as the `$\nu$-balancing
energy' \cite{Don3}, given by
\begin{equation} \label{Snu}  S_k(\omega_0, \phi) =\frac{kN_k}V \int_M \phi\; \nu^n - \log \det P, \end{equation}
where $\nu$ is a fixed volume form (independent of $\omega_0$) and $P$ is  positive Hermitian matrix, see \S \ref{FSECT} for the definition.

To put this choice of the action and the measure into context, we note that there are two basic approaches \cite{FKZ2} to choosing $S_k$
and $\dcal_k \phi$ in \eqref{MPI}. In the top-down approach, one starts with a Boltzmann weight $e^{-S(\phi)}$ and
path integral measure $\dcal \phi$ on $\kcalomega$, and uses the approximation of $\kcal_{\omega_0}$ by
$\bcal_k$ in order to regularize the formal continuous path integral over metrics \cite{BF,BFK}. In this case $k$ plays the role of a cut-off, and by sending $k\to\infty$ one should be able to recover the results in a given continuous theory. For instance,
one may try to approximate Liouville theory with an appropriate choice of $S_k, \dcal \phi_k$. 
  The second (``bottoms up") approach  is  to choose  $\dcal \phi_k$
and the  functionals
$S_k$ which naturally arise from the  symmetric space geometry of  $\bcal_k$ and from the geometric properties of Bergman metrics \cite{Don1}, and to investigate the asymptotics of integrals over $\bcal_k$ as $k \to \infty$.    Since $\bcal_k$
is a symmetric space, it  is natural to  choose the path integral measure $\dcal_k \phi$ to be Haar measure.   It is also
natural to choose $S_k$ to be a stability function or a functional closely related to stability. We review the notion of stability
in the following section \S \ref{STABSECT}, and refer to \cite{PS, Don2,Don3, T2, KLM, BGW} for several expositions of GIT
stability.

One would like the measures $e^{- S_k(\phi)} \dcal \phi_k$ to be probability measures, or at least finite measures
that can be normalized by dividing by \eqref{MPI}. But  Haar measure on  $\bcal_k$ is 
of infinite volume and in fact grows exponentially fast in terms of geodesic distance.  Thus, the first problem is to determine $\gamma_k$ 
so that  the integral \eqref{MPI} converges. This exponent depends on the choice of Haar measure for $\dcal \phi_k$
and also on the choice of $S_k$. The minimal $\gamma_k$ depends on the
growth of $S_k$ along  geodesics tending to infinity in $\bcal_k$.  The issue of the growth rate of geometrical
functionals along geodesics is precisely the stability problem \cite{PS3,PST,PST1}.

The existence of critical value of the coupling constant is akin to the existence of the $c=1$ barrier in Liouville theory. Recall \cite{KPZ} that the Lioville path integral over 2d metrics $g=e^\sigma g_0$ in the conformal class
\begin{equation}
\int e^{-\gamma S_L(g_0,\sigma)}\mathcal D \sigma,
\end{equation}
can be defined only for the values of the coupling constant $\gamma=\frac{26-c}{24\pi}\geq \gamma^{\rm min}$, where $\gamma^{\rm min}$ corresponds to the value of the central charge $c=1$. This bound can be interpreted as the effect of the non-trivial (non-flat) gravitational path integral measure \cite{D,DK}. Our analysis here suggests an explicit mechanism explaining how the bounds of this type may appear as a generic feature of the path integrals over metrics.

As mentioned above, in this article we choose the action functionals to be stability functions; 
the general notion is reviewed in  \S \ref{SF}, see Eq.\ \eqref{STABFUN}. Most of the important geometric functionals, such as
the Aubin-Yau, Mabuchi and Liouville energy, 
are stability functionals either on Bergman spaces $\bcal_k$ or on the space $\kcalomega$. 
 Stability functions are   asymptotically linear along geodesics and have an asymptotic
slope at infinity which measures stability along that geodesic.  We assume that
$(M, \omega_0)$ is in the stable case where the slope is positive for all geodesics of $\bcal_k$  for all $k$. If there
exists a csc (constant scalar curvature) metric in $\kcalomega$, then $(M, \omega_0)$ is in the stable case. In the stable case, 
$S_k$ is a $V$-shaped potential on $\bcal_k$, and    there exists a minimal value of the coupling constant $\gamma_k^{\rm crit}$, such that for $\gamma_k>\gamma_k^{\rm crit}$ \eqref{gammacrit} the integral \eqref{MPI} converges. This critical value defines a new stability invariant in \kahler geometry. The  linear growth rate of $S_k$ depends
on the choice of $S_k$ and also varies with  the geodesic. The integral stability index
$\gamma_k^{\rm crit}$ depends on the slope of $S_k$ along the geodesic ray  in the `worst' i.e. `least stable' direction. We then proceed to determine the exact $\gamma_k^{\rm crit}$ in the model of random Bergman metrics with action \eqref{Snu}.
The main result \eqref{pole} is that
\begin{equation} \label{MAINintro}\gamma_k^{\rm crit} = N_k -1. \end{equation}
For instance, in  the case of a Riemann surface of genus $h$,  $\gamma_k^{\rm crit} = k - h$.  
We also prove a more general version \eqref{poleep} in which we scale the eigenvalue directions of the symmetric space $\bcal_k$ by the weight $\epsilon_k$, which then becomes another parameter of the theory, along with $\gamma_k$. For the rescaled case $\gamma_k^{\rm crit}/\epsilon_k = N_k -1$. 

 Finally, we consider the problem of defining the large $k$ asymptotics. This can be done rigorously in the framework of  large deviations theory (we refer to \cite{ZZ} for
background and references in the context of this article). In large deviations theory,  one considers the sequence of probability measures on $\bcal_k$ induced by the integrals \eqref{MPI}. 
Since $\bcal_k \subset \kcalomega$, they may be regarded as a sequence of integrals over
$\kcalomega$ which are concentrated on $\bcal_k$. The purpose of large deviations theory
is to determine how the measures concentrate as $k \to \infty$.  In particular, one would
like to determine the asymptotic mass of random metrics in a fixed  Mabuchi-metric  ball of $\kcalomega$. The sequence of measures is said to satisfy an LDP  (large deviations principle)
with speed $N_k$ and rate function $I$ if $1/N_k^2\log$ of the probability measure tends to $I$
in a suitable sense. Proving that our measures \eqref{MPI} satisfy an LDP is one of the
ultimate goals of our work, but it is difficult to prove 
even for the simplest stability function (the $\nu$-balancing energy), and we only scratch
the surface in this article by studying upper bounds on the concentration of the measures.
Our  upper bound
 replaces  integration over spaces $\bcal_k$ of growing dimension by integration over a fixed limiting space of probability measures on $\mathbb R\times M$.  In the last section we determine, that for Riemann surfaces the large deviation principle holds if $\gamma_k/\epsilon_k$ is of order $N_k$ and $\gamma_k$ is of order one.

\section{\label{STABSECT} Random Bergman metrics and stability}

The purpose of this section is to review Bergman metrics and the notion of stability. In particular we introduce
the invariant $\gamma_k^{\rm crit}$ \eqref{gammacrit} and explain what it depends on. We should emphasize in advance
that the problem we pose in this section is very general, and that the integral with action \eqref{Snu} is only a simple
special case of the general stability integral problem. 

\subsection{Background on Bergman metrics}

Following the setup of \cite{FKZ2}, we consider the K\"ahler $(M,\omega)$ with the metric in a fixed K\"ahler class $\omega\in[\omega_0]$. The space of all K\"ahler metrics $\kcalomega$
on $M$ in the K\"ahler class $[\omega_0]$ is parametrized as 
\begin{equation}
\mathcal K_{\omega_0}= \{\phi\in C^{\infty}(M)/\mathbb R,\, \omega_\phi=\omega_0+i\p\bp\phi>0\},
\end{equation} 
Then there exists a  holomorphic  line bundle $L\to M$ with  hermitian metric $h_0$ whose curvature equals the background \kahler form $\omega_0=-i\p\bp\log h_0$. We then consider the tensor power $L^k$,  and choose a basis of sections $\{s_i(z)\}$, $i=1,\ldots,N_k$, such that it is orthonormal with respect to background metric
\begin{equation}\label{ONB}
\frac1V\int_M \bs_is_jh_0^k\,\omega_0^n=\delta_{ij},
\end{equation}
where $V=\int_M\omega^n$ is the volume of $(M,\omega_0)$, which is the same for all metrics $\omega\in\kcalomega$. The choice of the basis defines the Kodaira embedding $z \to [s_1(z), \dots, s_N(z)]$  of $M$ into the projective space $\mathbb{CP}^{N_k-1}$ where the sections live. We use this embedding in order to pull back the Fubini-Study metrics from $\mathbb{CP}^{N_k-1}$. Given the embedding associated to the orthonormal basis \eqref{ONB}, we obtain all others
using the action of $SL(N_k, \C)$ on $\CP^{N_k}$. In this way, we associate embeddings and Bergman metrics
to elements of $SL(N_k, \C)$. Since $SU(N_k)$ acts by isometries of the Fubini-Study metric, the resulting space of 
metrics is the quotient space of positive Hermitian matrices. Equivalently, to a positive Hermitian matrix $P$
we associate the metric
\begin{equation}
\label{BM}
\omega_{\phi(P)}=\frac1ki\p\bp\log \bs_i P_{ij}s_j,
\end{equation}
where summation over repeated indices is understood,
and in this way establish the identification
\begin{equation}
\label{bcalk}
\bcal_k \simeq SL(N_k,\mathbb C)/SU(N_k).
\end{equation} 
  Note that scalar multiplications of $P$ do not affect the Bergman metric, therefore we restrict the matrices $P$ from $GL(N_k,\mathbb C)/U(N_k)$ down to $SL(N_k,\mathbb C)/SU(N_k)$, by choosing the gauge: $\det P=1$.

\subsection{Distance and volume on $\bcal_k$}

The key fact \cite{T1,Z,C} is that, as $k\to\infty$, the space $\bcal_k $ becomes dense in the space of all K\"ahler metrics $\kcalomega$. This density is first of all point-wise: given any $\omega \in \kcalomega$, there exists a canonical
sequence $\omega_k \in \bcal_k$ so that $\omega_k \to \omega$ uniformly and with a complete asymptotic
expansion. Further, the global space $\bcal_k $ approximates $\kcalomega$ as Riemannian manifolds.
For instance, the geodesics of $\bcal_k$ tend to geodesics of $\kcalomega$. Moreover,  it has been shown by Chen-Sun \cite{CS}  that the distance function
$d_{\bcal_k}$ on $\bcal_k$ (with a proper normalization) tends to the distance function $d_M$ of the Mabuchi metric on $\kcal_{\omega}$. 

We are interested in the asymptotics of integrals over $\bcal_k$ as $k \to \infty$, and that requires the sequence
of symmetric space metrics on $\bcal_k$
 to be normalized properly. We recall that the symmetric space Riemannian metric on the space of positive
Hermitian matrices is invariant under the  $SL(N, \C)$ action, and is determined up to a  positive scalar multiple 
by the metric tensor at the identity element, which we refer to as the origin of the symmetric space. Thus,
the exponential map $\exp = \exp_I : T_I SL(N,\C) / SU(N) \to SL(N,\C) / SU(N), \; A \to e^A$ is a diffeomorphism.   At  the origin a tangent vector is a self-adjoint Hermitian
matrix $A$ whose norm  is a multiple of the Cartan-Killing norm $||A|| = \sqrt{\mbox{Tr} A^* A}$. Thus,  the distance from
$I$ to $e^A$ is $||A||$ (see \cite{H} for background on symmetric spaces).  One may choose any multiple
$c ||A||$ but temporarily we choose $c = 1$. In the next section we `scale' the metric by choosing a $k$-dependent
multiple $\epsilon_k $.

The canonical invariant (Haar) measure on the symmetric space \eqref{bcalk} reads
\begin{equation}
\label{measure}
\mathcal D_k\phi=\delta(\log\det P)[dP],\quad [dP]=\prod_{i<j}^{N_k}d\,\Im P_{ij}\,d\,\Re P_{ij}\;\prod_{i=1}^{N_k}dP_{ii},
\end{equation}
where the delta function constraint restricts the integration to $\bcal_k$. Using the  angular decomposition of positive-definite hermitian matrices
\begin{equation}
P=U^\dagger \Lambda U,
\end{equation}
we rewrite the measure explicitly as
\begin{equation}\label{CK}
[dP]=\Delta^2(\lambda)\,[d\lambda][dU],\quad\Delta(\lambda)=\prod_{i<j}(\lambda_i-\lambda_j)^2,\quad [d\lambda]=\prod_{i=1}^{N_k}d\lambda_i.
\end{equation}
We also write
\begin{equation}
\label{mui}
\Lambda_{ij}=\lambda_i\delta_{ij}=e^{\mu_i}\delta_{ij}, \quad \mu_i\in \mathbb R^{N_k},
\end{equation}
since the eigenvalues are positive. Then
\begin{equation}
\label{measure1}
\mathcal D_k\phi=\delta\left({\sum}_i\mu_i\right)\prod_{i<j}(e^{\mu_i}-e^{\mu_j})^2\prod_{i=1}^{N_k}d\mu_i\;[dU].
\end{equation}
The  integration measure over unitary matrices is normalized as
$$\int_{SU(N_k)}[dU]=1.$$

\subsection{\label{SCALINGSECT} Scaling}

We now scale the norm of a tangent vector $A$ at the identity element to $\epsilon_k ||A||$ and refer
to the corresponding  symmetric space metric on $\bcal_k$ by $g_{\epsilon_k}$. Thus, $g_1$ is the `standard' metric. 
There is no obvious way to fix the ambiguity in the one-parameter family of symmetric space metrics since
the volumes are infinite, unless we use the infinite dimensional geometry of $\kcalomega$ or the study
of the integrals \eqref{MPI}.

We now define dilation maps centered at the origin by
$$D_{\epsilon_k}: \bcal_k \to \bcal_k, \;\;  D_{\epsilon} (\exp A)  =  \exp \epsilon_k A . $$
The derivative of the dilation at the origin is the scaling map $A \to \epsilon_k A$. 
The pullback of the Cartan-Killing volume form \eqref{CK} under the dilation  is
\begin{equation}
\label{epsilondef}
\dcal^{\epsilon_k} \phi_k: =\delta(\epsilon_k\log\det P) [dP]_{\epsilon_k}=\delta\left(\epsilon_k{\sum}_i\mu_i\right)\prod_{i > j} (e^{\epsilon_k \mu_i} - e^{\epsilon_k \mu_j} )^2 \prod_{i=1}^{N_k}\epsilon_kd \mu_i\, [dU].
\end{equation}
The standard measure \eqref{CK} corresponds to $\epsilon_k = 1.$ 
The main point is that we can use this $\epsilon$-modified measure, with the definition of the Bergman metric \eqref{BM} kept intact. This way we can get different scaling limits, depending on the choice of $\epsilon_k$. We further write $Z^{\epsilon_k}_k(\gamma_k)$ for the partition function with the measure \eqref{epsilondef}.


As mentioned above, to obtain the Mabuchi metric $d_M$  on $\kcalomega$ as the limit
of the  symmetric space metric $d_{\bcal_k}$ on $\bcal_k$, one needs to rescale
the Cartan-Killing  metric  $g_{k}$  by the factor $\epsilon_k = k^{-1} N_k^{-1/2}$. As will be seen below,
other natural scalings arise when we consider the large $k$ limit of the integrals.

\subsection{\label{SF}Stability functions} 

Haar measure is a canonical choice of  measure, but it has infinite volume. To damp it out we use geometric functionals to define
Boltzmann weights \eqref{MPI}.  There are many potentially interesting functionals to employ.  The action functionals of eigenvalue type were considered before in \cite{FKZ4}. Here we 
choose $S$ to be a  geometric  functional.  In further work, we consider ball means. 
We first recall the definition of stability function and then specialize to the setting of $\bcal_k$.

The GIT stability problem involves a Hamiltonian action of a compact Lie group $G$  on a  symplectic manifold $(X, \omega)$,  with moment map $\Phi: X \to \mathfrak g^*$.  We will assume $G = SU(N)$ for simplicity.
 Let $G_{\C} = SL(N,\C)$ be the complexification of $G$. A
stability function $\hat{\psi}_v$  is a function on $G_{\C}$  whose gradient is the moment map.  The  $SL(N,\C)$ action lifts
from $X$ to the space $H^0(X, L)$ of holomorphic sections of a  Hermitian line bundle   $L \to X$. Let
$v  \in H^0(X, L)$. Then the stability function associated to $v$ is defined by
\begin{equation} \label{STABFUN} \hat{\psi}_v(g) = \log ||g \cdot v||^2, \;\;\; g \in G_{\C}. \end{equation}
The norm $||\cdot||$ is assumed to be invariant under  $SU(N)$ and so one restricts to the imaginary complex directions
$\exp \sqrt{-1} \mathfrak g \cdot v. $  The stability integrals have the form,
$$\int_{\exp \sqrt{-1} \mathfrak g \cdot v} f  e^{\lambda \psi_v} dV, \;\;\; f \in C_b(G_{\C}) $$
where $dV$ is the invariant measure on $
\exp \sqrt{-1} \mathfrak g $ and $f$ is a bounded  continuous function.  As will be reviewed below, stability functions $\psi_v$ are asymptotically linear, geodesically
convex functions. When restricted to a geodesic in $G_{\C}$, i.e. a one-parameter subgroup, the stability function
has a tangent line at infinity, whose slope is known as the asymptotic slope. For convergence of the integral, the $y$-intercept
is also important.  Aside from this article, they are studied in \cite{BGW} but only
in the case where  $f $ is compactly supported.  For the partition function, $f \equiv 1$ and the difficult problem is
to determine the $\lambda$ for which the integral converges. We refer to \cite{DK,PS,W,KLM} for background on stability 
functions.

Stability functions are central to the
question, when do canonical metrics (i.e. K\"ahler-Einstein metrics or more general metrics of constant scalar curvature)
exist 
in  $\kcalomega$? As is well-known,  a unique constant scalar curvature (csc) metric exists on any  Riemann surface in a given conformal or, equivalently, K\"ahler class. On \kahler manifolds of complex dimension $n$, this is not always the case.  The Yau-Tian-Donaldson program  is to prove that the existence of
canonical metrics is equivalent to the properness of a special stability function known as the Mabuchi energy.  
   Roughly speaking, stability
means that the asymptotic slopes of the relevant stability functions (known as the Mabuchi K-energy) are positive
in all directions.
The program has recently been brought to a successful conclusion in the case of K\"ahler-Einstein metrics in
 works of X.~Chen, S.~K.~Donaldson and S.~Sun and in works of G.~Tian, but it would take us
too far afield to describe those developments; for purposes of this article it is sufficient to cite  \cite{Th,PS,T3} for surveys and background.

In this article, we only consider $(M, \omega)$ for which   the stability condition is satisfied, i.e. when $\kcalomega$
is known to contain a csc metric. Thus the stability functions
in our setting are
$V$-shaped
potential and for sufficiently large $\lambda$ the stability integral should converge.  
 Examples of stability functions in the setting of Bergman metrics include Aubin-Yau, Mabuchi and Liouville functionals,
which are defined in the next section \S \ref{FSECT}. When restricted to Bergman metrics, these functionals are bounded from below and exhibit universal linear behavior \cite{PST,PST1,PS1,PS3,PS5} at the geodesic infinity.

 We now explain this in detail. As in \eqref{bcalk}, a geodesic of $\bcal_k$ corresponds to a one parameter
subgroup of $SL(N_k, \C)$ and its initial direction is defined by a pair $(U, \Lambda)$ where $U \in U(N_k)$
and $\Lambda$ is a positive diagonal $N_k \times N_k$ matrix.   Geodesic coordinates on $\bcal_k$ are introduced by introducing a radial coordinate $t$ in the euclidean space of $\mu_i$ \eqref{mui} as follows
\begin{equation}
\lambda_i=e^{\mu_i}=e^{ta_i},\quad\sum_{i=1}^{N_k}a^2_i=1,\quad t\in[0,\infty).
\end{equation}
In this coordinate system the geodesic infinity corresponds to $t\to\infty$. Each one-parameter geodesic is thus parameterized as 
\begin{equation}
\label{oneparam}
P_t=U^\dagger e^{ta}U.
\end{equation}

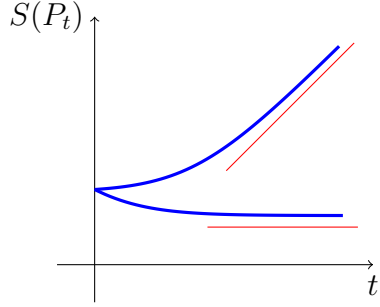
\begin{figure}[h]
\begin{center}
\begin{tikzpicture}
\draw[->] (-.5,0) -- (3.7,0) node[below] {$t$};
  \draw[->] (0,-.5) -- (0,3.3) node[left] {$S(P_t)$};
  \draw[scale=0.5,domain=0:6.5,smooth,variable=\x,blue,very thick] plot ({\x},{2+ln(14/15+exp(\x)/15)});
  \draw[scale=0.5,domain=2.5:5.9,smooth,variable=\y,red]  plot ({\y+1},{\y}) ;
\draw[scale=0.5,domain=0:6.6,smooth,variable=\x,blue, very thick] plot ({\x},{2+ln(1/2+exp(-\x)/2)});
  \draw[scale=0.5,domain=2:6,smooth,variable=\y,red]  plot ({\y+1},{1});
\end{tikzpicture}
\end{center}
\caption{Graphs of stability function along geodesics}\protect\label{fig:fig1}
\end{figure}

The restriction of a stability function $S$ to a geodesic $P_t$ defines a function $S (P_t) $ on $\R$ which  is asymptotically
linear in the sense that 
\begin{equation}
\label{asymptotics}
S(P_t)\sim At-B, \;\; \mbox{as}\; t \to \infty,
\end{equation}
where $A$ and $B$ are the constants depending on $(U, \Lambda)$, called the asymptotic slope and the $y$-intercept, respectively.  Assuming $S$ is bounded from below, two kinds of behavior at infinity are possible, see Fig.\ \ref{fig:fig1}, depending on whether $A>0$ for all geodesics, or there are some geodesics with $A=0$. In the first, stable case the corresponding functional obviously has a critical point somewhere inside $\bcal_k$. In the second, unstable case the critical point is not achieved at finite $t$.

The asymptotic slope $A$ appears in the  general context of stability functions in GIT (see \cite{KLM,W}).  The articles \cite{L,Zh} contain
perhaps the earliest studies of asymptotics slopes for functionals of  Bergman metrics. 
The asymptotic slope $A$ is calculated explicitly in the case of Riemann surfaces for several energy functionals
in \cite{PST,PS3,PS}.  

As mentioned above, we assume we are in the stable case, so that the
asymptotic linear growth  of the action \eqref{asymptotics} damps out non-compact directions in $\bcal_k$ in the integral \eqref{MPI}. On the other hand, the Vandermonde determinant blows up along non-compact directions. Indeed, in geodesic coordinates the volume form \eqref{measure1}  for the standard Cartan-Killing metric reads
\begin{eqnarray}
\nonumber
\mathcal D_k\phi=\delta\left({\sum}_ia_i\right)\prod_{i<j}(e^{ta_i}-e^{ta_j})^2
 t^{N_k-2}dt[d\Omega][dU],
\end{eqnarray}
where $[d\Omega]$ denotes the volume form on the unit sphere $S^{N_k-1}$ in the $\mu$-space.
We can determine its asymptotic behavior if we adopt the ordering 
\begin{equation}
\label{sum}
\mu_1\leqslant \mu_2\leqslant\ldots\leqslant \mu_{N_k}, 
\end{equation}
at the expense of an overall factor $N_k!$ in the partition function. This condition restricts the integration domain to a sector of the sphere $S^{N_k-1}$, cut out by \eqref{sum}. The integration domain is restricted further to the intersection of this sector with the hyperplane in $\mathbb R^{N_k}$, defined by the delta-function constraint.
Then along the non-compact directions the measure tends to
\begin{equation}
\label{asymp2}
\mathcal D_k\phi\simeq \delta\left({\sum}_ia_i\right)e^{2t\sum_{j=1}^{N_k}ja_j} t^{N_k-2}dtd\Omega[dU],\quad{\rm as}\; t\to\infty
\end{equation}
and one can check that the sum in the exponent is always non-negative, leading to the growth of the measure at geodesic infinity.

For the $\epsilon_k$-deformed volume form \eqref{epsilondef} the analogous calculation gives 
\begin{equation}
\label{asymp2sc}
\mathcal D^{\epsilon_k}_k\phi\simeq \delta\left({\sum}_ia_i\right)e^{2t \epsilon_k \sum_{j=1}^{N_k}ja_j} t^{N_k-2}dtd\Omega[dU],\quad{\rm as}\; t\to\infty.
\end{equation}
 
From the asymptotics \eqref{asymptotics} and \eqref{asymp2} it follows that there is a competition at geodesic infinity between the Boltzmann weight $e^{-\gamma_kS_k}$ and the path integral measure. Therefore the partition function may not converge if the coupling constant is not sufficiently large. This leads us to define the new {\it stability invariant},  as the following critical value of the coupling constant
\begin{equation}\label{gammacrit}
\gamma_k^{\rm crit}=\inf\{\gamma_k\;|\; Z_k(\gamma_k)<\infty\}.
\end{equation}
In other words, for the coupling constants less than the critical the partition function ceases to converge, even at finite $k$. The critical value depends in principle on the \kahler manifold $(M,[\omega_0])$, as well as on the action functional. In the stable case $\gamma_k^{\rm crit}<\infty$. Intuitively, it is clear from Fig.\ \ref{fig:fig1}, that geodesic directions with slow growth at infinity push the critical gamma upwards, since these "bad directions" worsen the convergence of the partition function. Therefore  $\gamma_k^{\rm crit}$ measures the minimal slope and the size of the set of the worst directions for a given stability function. 

For the partition function with the $\epsilon_k$-deformed measure \eqref{epsilondef} the critical coupling constant $\gamma_k$ depends on the scaling parameter $\epsilon_k$. In this case we write
\begin{equation}\label{gammacritep}
\gamma_k^{\epsilon_k, \rm crit}=\inf\{\gamma_k\;|\; Z_k^{\epsilon_k}(\gamma_k)<\infty\},
\end{equation}
and the case \eqref{gammacrit} corresponds to
$\epsilon_k = 1$.

\section{\label{FSECT} $S_\nu$ functional, balancing energy and Aubin-Yau action}

\subsection{Balanced metrics}

We would like to choose  actions $S_k$ on $\bcal_k$ whose critical points are closely related to constant scalar curvature \kahler metrics on $M$. The analog  in $\bcal_k$ of the csc metric in $\kcalomega$  is the so-called balanced metric \cite{L,Don2}. The balancing condition is given by the following integral equation
\begin{equation}
\label{bal1}
\frac{N_k}V\int_M\frac{\bs_is_j}{\bs_l P_{lm}s_m}\omega^n_{\phi(P)}={P^{-1}}_{ji}.
\end{equation}
If  this equation is solved by a matrix $P=P_0$, then the corresponding Bergman metric $\omega_{\phi(P_0)}$ is called the balanced metric. The name "balanced" refers to the natural interpretation this metric has in terms of Kodaira embedding. Note that the integral in \eqref{bal1} is the moment of inertia matrix for the embedding.  If we write $P_0=A_0^\dagger A_0$, for $A_0\in GL(N_k,\mathbb C)$, then for the embedding $M\to (A_0s)_i(z)\in\mathbb{CP}^{N_k-1}$ the moment of inertia matrix is diagonal, i.e. one can say that the image of $M$ is "balanced" inside the projective space.

If  there exists csc \kahler metric on $(M,\omega_0)$, then there is a unique balanced metric for each large enough level $k$, and  as $k\to\infty$ the sequence of balanced metrics converges to the csc metric on $M$ \cite{Don2}.  This situation corresponds to the upper graph in Fig.\ \ref{fig:fig1}. 

Due to a highly nontrivial dependence of the integrand in Eq.\ \eqref{bal1} on $P$, this equation is in general hard to solve,
and balanced metrics are rarely known explicitly  except for special cases such as the complex projective spaces $ \CP^m$ \cite{Don3} or abelian varieties \cite{Wang}. Fortunately, there exists another, easier, sequence of metrics, called $\nu$-balanced \cite{Don3}, with similar properties. This metric is defined with respect to fixed volume form $\nu^n(z)$ on $M$. One has to solve a similar integral equation \eqref{bal1}, but now with the fixed volume form
\begin{equation}
\label{bal2}
\frac{N_k}V\int_M\frac{\bs_is_j}{\bs_l P_{lm}s_m}\nu^n(z)={P^{-1}}_{ji}.
\end{equation}
It is shown in \cite{Don3}, that the sequence of $\nu$-balanced metrics converges as $k\to\infty$ to the \kahler metric with the volume $\nu^n$ \cite{Don3}. In particular, for the Calabi-Yau (Ricci-flat) manifolds one can choose $\nu^n=(-i)^n\theta\wedge\theta$, where $\theta$ is the holomorphic $n$-form.

So far we did not make any choice for the background metric $\omega_0$. The simplest choice for $\omega_0$ is the balanced metric itself, as defined by Eq.\ \eqref{bal1}. This is equivalent to saying, that \eqref{bal1} is satisfied for the identity matrix $P_{ij}=\delta_{ij}$. With this choice of the background metric, and thus the basis of sections $s_i$, the Bergman metrics are written as
\begin{equation}
\label{assump}
\omega_{\phi(P)}=\omega_0+i\p\bp\phi(P)=\frac1ki\p\bp\log\sum_l|s_l|^2+ \frac1k i\p\bp\log \frac{\bs_iP_{ij}s_j}{\sum_l|s_l|^2}.
\end{equation}
We also introduce the \kahler potential for the Bergman metric 
\begin{equation}
\phi(P)= \frac1k\log \frac{\bs_iP_{ij}s_j}{\sum_l|s_l|^2}.
\end{equation}
In particular, at the balanced metric $P_{ij}=\delta_{ij}$ the \kahler potential is zero. The 
factor  $\frac{1}{k}$ is  a standard normalization which  makes the family of Bergman  potentials uniformly bounded in a certain sense, see \eqref{UNIFBOUND}.

\subsection{Actions}

Now we are ready to construct the action functionals on $\bcal_k$, which reproduce Eq. \eqref{bal1} and \eqref{bal2} as critical points,
\begin{eqnarray}
\label{critp}
\delta S_k(\omega_0,\phi(P))&=&\left(\frac{N_k}V\int_M\frac{\bs_is_j}{\bs_l P_{lm}s_m}\omega^n_{\phi(P)}-{P^{-1}}_{ji}\right)\delta P_{ij},\\
\delta S_{\nu,k}(\omega_0,\phi(P))&=&\left(\frac{N_k}V\int_M\frac{\bs_is_j}{\bs_l P_{lm}s_m}\nu^n-{P^{-1}}_{ji}\right)\delta P_{ij}.
\end{eqnarray}
The subscript $k$ indicates that the functionals are defined on $\bcal_k$. The additional subscript $\nu$ for the second functional emphasizes, that apart from the background metric $\omega_0$ it also depends on the choice of the background volume form $\nu^n$.
The explicit formulas for the above action functionals read
\begin{eqnarray}
\label{AY1}
S_k(\omega_0,\phi(P))&=&\frac{kN_k}{(n+1)V}\int\phi(P)\sum_{p=0}^n\omega_{\phi(P)}^p\wedge\omega_0^{n-p}-\log\det P,\\\label{AY}
S_{\nu,k}(\omega_0,\phi(P))&=&\frac{kN_k}V\int_M\phi(P)\,\nu^n-\log\det P.
\end{eqnarray}
We stress, that despite that the second functional looks very easy in terms of $\phi$, it is nevertheless a non-trivial functional on $P$, and the $\nu$-balanced metric is its unique global minimum on $\bcal_k$.
Up to the overall normalization factor $kN_k$, the first functional above is the normalized restriction to $\bcal_k$
of the  Aubin-Yau functional,
\begin{equation}
S_{AY}(\omega_0,\phi)=\frac1{(n+1)V}\int\phi\sum_{p=0}^n\omega_{\phi}^p\wedge\omega_0^{n-p},
\end{equation}
on $\kcalomega$. (The term `Aubin-Yau' functional is not standard and is adopted from \cite{PS3}.) The restricted functional   is also called the balancing energy, it was studied in detail in \cite{Fine}. 
The second functional, which may be called $\nu$-balancing energy, is the normalized  restriction  to $\bcal_k$
of the  following simple functional on $\kcalomega$, 
\begin{equation}
S_{\nu}(\omega_0,\phi)=\frac1V\int_M\phi\,\nu^n,
\end{equation}
and is discussed  also  in \cite{Don3}.  




Both functionals in \eqref{AY} are convex along  one-parameter geodesics \eqref{oneparam} on $\bcal_k$. For the Aubin-Yau functional restricted to  $\bcal_k$ this was shown in \cite{Don4}. The the functional $S_\nu$ the calculation is straightforward. Its second derivative is explicitly non-negative
\begin{equation}
\label{convex}
\frac{d^2}{dt^2}S_{\nu,k}(\omega_0,\phi(P_t))=\frac{kN_k}V\int_M\ddot{\phi}(P_t)\nu^n=\frac{N_k}{2V}\sum_{i,j}(a_i-a_j)^2\int_M\rho_i(t)\rho_j(t)\, \nu^n\geq0
\end{equation}
where 
\begin{equation}
\rho_i(t)=\frac{(\bs U^\dagger)_ie^{ta_i}(Us)_i}{\bs P_t s}
\end{equation}
is a non-negative definite density. The functionals \eqref{AY} are normalized such that at the critical points \eqref{critp} their values are exactly zero. Hence, they are bounded from below by zero.

For a particular choice of the background volume form $\nu^n=\omega_0^n$ the Aubin-Yau and $S_\nu$-functional are related by the following formula
\begin{equation}
F_{AY}(\omega_0,\phi)=S_{\omega_0}(\omega_0,\phi)-\frac1{(n+1)V}\int_M i\p\phi\wedge\bp\phi\wedge\sum_{p=0}^{n-1}(p+1)\omega_0^p\wedge\omega_\phi^{n-1-p},
\end{equation}
see e.g. \cite{CT}.
It follows that the lower bound for $F_k$
\begin{equation}
F_k(\omega_0,\phi(P))\geq0
\end{equation}
implies the lower bound for $S_{\omega_0,k}$
\begin{equation}
S_{\omega_0,k}(\omega_0,\phi(P))\geq\frac{kN_k}{(n+1)V}\int_M i\p\phi\wedge\bp\phi\wedge\sum_{p=0}^{n-1}(p+1)\omega_0^p\wedge\omega_\phi^{n-1-p}\geq0.
\end{equation}
These formulas become more transparent, when restricted to complex dimension $n=1$. In this case $M$ is a compact Riemann surface of genus $g$ and the number of sections is 
\begin{equation}
N_k=k+1-g.
\end{equation}
Then 
\begin{equation}
S_{AY}(\omega_0,\phi)=\frac1A\int_M\frac12\phi\, i\p\bp\phi+\phi\omega_0, 
\end{equation}
is bounded from below by zero, and it follows that
\begin{equation}
S_{\omega_0,k}(\omega_0,\phi(P))\geqslant\frac{kN_k}{2A}\int_M i\p\phi\wedge\bp\phi=\frac{kN_k}{2A}\int_M |\p\phi|_{\omega_0}^2\omega_0
\end{equation}

In the next section we determine the exact value of the stability invariant $\gamma_k^{\rm crit}$ for the $S_\nu$ functional.

\section{Asymptotic slopes and y-intercepts of $\nu$-balancing energy }

\subsection{Asymptotic slope}

Our goal now is to determine $\gamma_k$ for which the integral \eqref{MPI} converges when $S_k =
S_{\nu,k}$ is $\nu$-balancing energy. 
To this end we find the asymptotics of the $\nu$-balancing energy along the one-parameter geodesics \eqref{oneparam}. Taking into account the delta-function constraint, we have
\begin{eqnarray}
S_{\nu,k}(\omega_0,\phi(P_t))&=&\frac{kN_k}V\int_M\frac1k\log \frac{\bs_i(U^\dagger e^{ta}U)_{ij}s_j}{\sum_l|s_l|^2}\; \nu^n-\log\det U^\dagger e^{ta}U=\\
&=&N_ka_{N_k}t+\frac{N_k}V\int_M\log \frac{\sum_i(\bs U^\dagger)_i e^{t(a_i-a_{N_k})}(Us)_i}{\sum_l|s_l|^2}\; \nu^n.
\end{eqnarray}
Due to the exponential suppression at large $t$, for the second term we get
\begin{equation}
\label{asymp}
S_{\nu,k}(\omega_0,\phi(P_t))=N_ka_{N_k}t+\frac{N_k}V\int_M\log \frac{\sum_{i=N_k-r}^{N_k}(\bs U^\dagger)_i (Us)_i}{\sum_l|s_l|^2}\; \nu^n,\quad {\rm as}\;t\to\infty
\end{equation}
where $r\leqslant N_k-1$ accounts for possible degeneracy  of the highest eigenvalue $a_{N_k}$. In the generic situation $r=0$, and only the section $(Us)_{N_k}$ remains. 

This is exactly the linear asymptotics of Eq.\ \eqref{asymptotics}. Therefore  the asymptotic slope of the tangent line is $A=N_ka_{\rm max}$. We see that the slope is positive, therefore the functional is convex along geodesics with positive asymptotic slopes, so that the Boltzmann weight $e^{-\gamma_kS_{\nu,k}(\omega_0,\phi(P))}$ indeed serves as a damping factor for the measure. The $y$-intercept is given by the second term in \eqref{asymp}. We need to show, that the $y$-intercept is uniformly bounded from below, for all geodesics. For this we need a uniform bound of the form,
\begin{equation} \label{UBLOG} \frac{1}{k} \left| \frac1V\int_M  \log |s(z)|^2 \omega_0 \right|  \leq C\end{equation}
for the family of  $L^2$-normalized holomorphic sections
$$s \in H^0(M, L^k): ||s||_{L^2} = 1,$$   When there are repeated eigenvalues, we may have to sum up to as many as $N_k -1$
eigenvalues.   The family
\begin{equation} \label{FAMILY} \left\{\frac{1}{k} \log |s_k(z)|^2_{h^k}\right\} \end{equation}
is $\omega_0$-plurisubharmonic and uniformly bounded above. In fact, there exists a constant $C > 0$
so that $\frac{1}{k} \log |s_k(z)|^2_{h^k} \leq C \frac{\log k}{k}$, as one sees using the reproducing formula
$s_k  = \Pi_ks_k$ and the fact that such bounds hold for $\Pi_k$  where $\Pi_k$ is the Szeg\"o kernel \cite{C,Z}. Here,  $\omega_0$-plurisubharmonic means that 
$\frac{1}{k}\log |s_k(z)|^2_{h^k}$ is locally the sum of a subharmonic function plus a smooth potential for $\omega_0$. There is a general
result regarding families of subharmonic functions which are uniformly bounded above which also applies to
families of bounded $\omega_0$-plurisubharmonic functions (see
\cite[Theorem~4.1.9]{Ho}):
\medskip

{\it The set  $SH(X)$ of $\omega_0$-subharmonic functions on the compact manifold  $M$ is closed
in the $L^1$ topology. Let $\{v_j\}  $ be a family of $\omega_0$-subharmonic functions on $M$ which has a uniform upper bound. Then either $v_j \to -\infty$ uniformly on every
compact set, or else there exists a subsequence $v_{j_k}$ which 
converges in $L^1(M)$ to some $\omega_0$-subharmonic function  $v \in L^1(M)$. Further, $ \limsup_j
v_j(x) \leq v(x) \;\mbox{ with equality almost everywhere} $.
}

It is straightforward to check that no $L^2$-normalized sequence from \eqref{FAMILY} tends to $- \infty$ uniformly on $M$,
since it contradicts that the $L^2$-norms equal one. Therefore the $L^1$ norms of the family \eqref{FAMILY} form
a bounded set, i.e. there exists $C > 0$ (independent of $k$) so that
\begin{equation} \label{UNIFBOUND} \frac1V\int_M \frac{1}{k} \left| \log |s_k(z)|^2_{h^k} \right| \nu^n  \leq C. \end{equation}
It follows, that the $y$-intercept is uniformly bounded. 
\begin{equation}
\frac{N_k}V\int_M\log \frac{\sum_{i=N_k-r}^{N_k}(\bs U^\dagger)_i (Us)_i}{\sum_l|s_l|^2}\geq -CkN_k,
\end{equation}
where $C$ is $k$-independent constant.

\subsection{Other geometric functionals}

The asymptotic slope $A$ is calculated explicitly in the case of Aubin-Yau energy and Mabuchi energy on  Riemann surfaces 
in \cite{PST,PS3}.  For the higher dimensional case see \cite{PST1}.  Unlike the $\nu$-balancing energy, the asymptotic
slope of the Aubin-Yau and Mabuchi functionals  is non-constant  and there is no lower bound for the $y$-intercept, which may tend to $- \infty$ as the
geodesic ray tends to a limit direction where the asymptotic slope jumps down.  Hence the analysis of integrals
\eqref{MPI} becomes much more complicated when the action is the Aubin-Yau energy or other stability functions. 

There is an alternative approach to the analysis of $\gamma_k^{\rm crit}$ which we plan to discuss elsewhere. 
It is based on  lower bounds on the Hessian
of the Aubin-Yau energy near the balanced metric, which give reasonably sharp  bounds on $\gamma_k^{\rm crit}$
for that functional. In effect, we use the tangent line at a fixed distance from the origin instead of using the tangent
line at infinity. 
One of our motivations to study the balancing energy  is that we can find exact values of the asymptotic slopes,
$y$-intercepts and $\gamma_k^{\rm crit}$,  which can be compared with the bounds we get using Hessian estimates.

To our knowledge, the minimal asymptotic slope of the principal geometric functionals (Mabuchi energy, Aubin-Yau energy) 
along geodesics of $\bcal_k$ is not known at this time. Nor are the geodesics on which the functional has its minimal 
slope. These rays are stable analogues of the `optimal test configurations' of \cite{Sz}, where the `worst' rays
are determined in the unstable case for toric varieties.

\section{Exact bound on the coupling constant}

\subsection{Critical $\gamma_k$ for the standard measure}
Now we consider the partition function \eqref{MPI} with the action given by the functional $S_{\nu,k}$,
\begin{equation}
\label{IPI}
Z_k(\omega_0,\nu;\gamma_k)=\int_{\bcal_k}e^{-\gamma_kS_{\nu,k}(\omega_0,\phi(P))}\mathcal D_k\phi
\end{equation}
where $D_k\phi$ is the invariant measure \eqref{measure} on $\bcal_k$ without scaling. We then modify the
calculation for the $\epsilon_k$-deformed volume form.
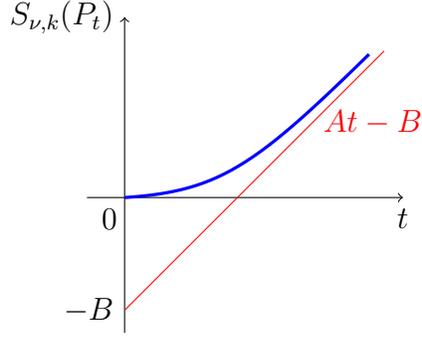
\begin{figure}[h]
\begin{center}
\begin{tikzpicture}
\draw[->] (-.5,0) -- (3.7,0) node[below] {$t$};
\draw (-.2,0) node[below]{$0$};
\draw (0,-1.5) node[left]{$-B$};
  \draw[->] (0,-1.8) -- (0,2.4) node[left] {$S_{\nu,k}(P_t)$};
  \draw[scale=0.5,domain=0:6.5,smooth,variable=\x,blue, very thick] plot ({\x},{ln(14/15+exp(\x)/15)});
  \draw[scale=0.5,domain=-3:3.9,smooth,variable=\y,red]  plot ({\y+3},{\y});
\draw[red] (2.5,1) node[right]{$At-B$};
\end{tikzpicture}
\end{center}
\caption{Graph of $S_{\nu,k}$ functional is above its tangent line at infinity}
\protect\label{fig:fig2}
\end{figure}
Taking into account convexity \eqref{convex}, we can bound the $S_{\nu,k}$ functional from below by its tangent line at infinity \eqref{asymp}, as shown in Fig.\ \eqref{fig:fig2}. This provides the upper bound on the partition function
\begin{eqnarray}
\label{boundint}
\nonumber
&&Z_k(\nu,\omega_0;\gamma_k)=\\\nonumber&&= N_k!\int_{SU(N_k)}\int_{\mu_1\leqslant\ldots\leqslant\mu_{N_k}} e^{-\gamma_kS_{\nu,k}(\omega_0,\phi(U^\dagger e^\mu U))} \delta\left({\sum}_i\mu_i\right)\Delta^2(e^\mu)\prod_{i=1}^{N_k}d\mu_i\;[dU]\\\label{upperb}
&&\leqslant N_k!\int_{\mu_1\leqslant\ldots\leqslant\mu_{N_k}} e^{-\gamma_k(N_k\mu_{N_k}-CkN_k)} \delta\left({\sum}_i\mu_i\right) \Delta^2(e^\mu)\prod_{i=1}^{N_k}d\mu_i\\\label{lastint}&&\leqslant
e^{kN_k\gamma_kC}N_k!\int_{\mu_1\leqslant\ldots\leqslant\mu_{N_k}} e^{-\gamma_kN_k\mu_{N_k}} \delta\left({\sum}_i\mu_i\right) e^{2\sum_{j=1}^{N_k}j\mu_j} \prod_{i=1}^{N_k}d\mu_i.
\end{eqnarray}
Note that in the third line we singled out the steepest direction in the Vandermonde determinant, which controls the critical gamma. Therefore we now only need to determine for which $\gamma_k$ the last integral \eqref{lastint} converges. This can be done by routinely integrating out all $\mu_i$ starting from $\mu_1$ up to $\mu_{N_k}$ while keeping only the pieces with the worst divergence at each step. As a check, in the next section we evaluate the integral in the second line explicitly and confirm that $\gamma_k^{\rm crit}$ we get this way is indeed exact. However, the advantage of the routine method is that it also works for more general asymptotics \eqref{asymptotics}. We get
\begin{eqnarray}
&&\nonumber\int_{\mu_1\leqslant\ldots\leqslant\mu_{N_k}} e^{-\gamma_kN_k\mu_{N_k}} \delta\left({\sum}_i\mu_i\right) e^{2\sum_{j=1}^{N_k}j\mu_j} \prod_{i=1}^{N_k}d\mu_i=\\&&\nonumber=
\nonumber\int_{-\frac12\sum_{j=3}^{N_k}\mu_j}^{\mu_3}d\mu_2
\int_{-\frac13\sum_{j=4}^{N_k}\mu_j}^{\mu_4}d\mu_3\ldots
\int_{-\frac1{N_k-1}\mu_{N_k}}^{\mu_{N_k}}d\mu_{N_k-1}\int_0^\infty d\mu_{N_k}
\;e^{-\gamma_kN_k\mu_{N_k}+2\sum_{j=2}^{N_k}(j-1)\mu_j}\\
&&\nonumber\leqslant
\prod_{l=2}^{m-1}\frac1{l(l-1)}\int_{-\frac1m\sum_{j=m+1}^{N_k}\mu_j}^{\mu_{m+1}}d\mu_m
\ldots
\int_{-\frac1{N_k-1}\mu_{N_k}}^{\mu_{N_k}}d\mu_{N_k-1}\int_0^\infty d\mu_{N_k}\\\nonumber&&
e^{-\gamma_kN_k\mu_{N_k}+2\sum_{j=m+1}^{N_k}(j-1)\mu_j+m(m-1)\mu_m}\\\nonumber&&\leqslant \prod_{l=2}^{N_k-1}\frac1{l(l-1)}\int_0^\infty d\mu_{N_k}e^{-\gamma_kN_k\mu_{N_k}+N_k(N_k-1)\mu_{N_k}}=\frac1{N_k(\gamma_k-N_k+1)}\cdot\prod_{l=2}^{N_k-1}\frac1{l(l-1)}.
\end{eqnarray}

The last integral, and therefore the full partition function, converges if 
\begin{equation}
\label{pole}
\gamma_k>\gamma_k^{\rm crit}=N_k-1.
\end{equation}
In complex dimension $n$, $N_k$ grows as $k^n+\frac12c_1(M)k^{n-1}$. In complex dimension one we have
\begin{equation}
\gamma_k^{\rm crit}=k-h,
\end{equation}
where $h$ is the genus of a Riemann surface.

\subsection{Scaling by $\epsilon_k$}

Now we perform the same calculation for the measure \eqref{epsilondef}. The
calculations change in the following obvious way
\begin{eqnarray}
\label{boundintep}
\nonumber
&&Z_k^{\epsilon_k}(\nu,\omega_0;\gamma_k)=\\\nonumber&&= N_k!\int_{SU(N_k)}\int_{\mu_1\leqslant\ldots\leqslant\mu_{N_k}} e^{-\gamma_kS_{\nu,k}(\omega_0,\phi(U^\dagger e^{ \mu} U))} \delta\left(\epsilon_k{\sum}_i\mu_i\right)\Delta^2(e^{\epsilon_k\mu})
\prod_{i=1}^{N_k}\epsilon_kd\mu_i\;[dU]\\\nonumber&&= N_k!\int_{SU(N_k)}\int_{\mu_1\leqslant\ldots\leqslant\mu_{N_k}} e^{-\gamma_kS_{\nu,k}(\omega_0,\phi(U^\dagger e^{ \mu/\epsilon_k} U))} \delta\left({\sum}_i\mu_i\right)\Delta^2(e^\mu)\prod_{i=1}^{N_k}d\mu_i\;[dU]\\\label{upperb}
&&\leqslant N_k!\int_{\mu_1\leqslant\ldots\leqslant\mu_{N_k}} e^{-\gamma_k(N_k \mu_{N_k}/\epsilon_k-CkN_k)} \delta\left({\sum}_i\mu_i\right) \Delta^2(e^\mu)\prod_{i=1}^{N_k}d\mu_i\\\label{lastint}&&\leqslant
e^{kN_k\gamma_kC}N_k!\int_{\mu_1\leqslant\ldots\leqslant\mu_{N_k}} e^{-\gamma_kN_k \mu_{N_k}/\epsilon_k} \delta\left({\sum}_i\mu_i\right) e^{2\sum_{j=1}^{N_k}j\mu_j} \prod_{i=1}^{N_k}d\mu_i\\ \nonumber &&\leqslant\frac{1}{N_k(\gamma_k/\epsilon_k-N_k+1)}\cdot\prod_{l=2}^{N_k-1}\frac1{l(l-1)}.
\end{eqnarray}

We see that the $\epsilon_k$ scaling just changes $\gamma_k \to \gamma_k/\epsilon_k$ in the integral over diagonal matrices; the integral over $U(N_k)$ is not involved in the upper bound.
Therefore, the scaled partition function  converges if 
\begin{equation}
\label{poleep}
\gamma_k>\gamma_k^{\epsilon_k, \rm crit}=\epsilon_k (N_k-1).
\end{equation}
The lower bound similarly changes only by changing $\gamma_k \to \epsilon_k^{-1} \gamma_k$, and by the
analysis in the unscaled case we see that \eqref{poleep} is sharp.

\subsection{Upper and lower bounds on the partition function}
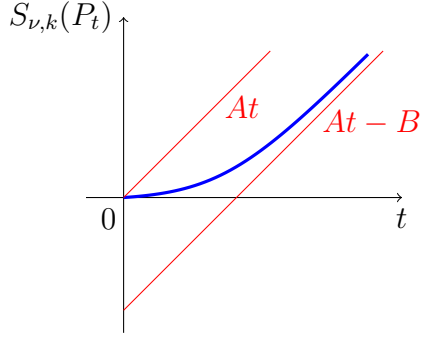
\begin{figure}[h]
\begin{center}
\begin{tikzpicture}
\draw[->] (-.5,0) -- (3.7,0) node[below] {$t$};
\draw (-.2,0) node[below]{$0$};
  \draw[->] (0,-1.8) -- (0,2.4) node[left] {$S_{\nu,k}(P_t)$};
  \draw[scale=0.5,domain=0:6.5,smooth,variable=\x,blue, very thick] plot ({\x},{ln(14/15+exp(\x)/15)});
  \draw[scale=0.5,domain=-3:3.9,smooth,variable=\y,red]  plot ({\y+3},{\y});
  \draw[scale=0.5,domain=0:3.9,smooth,variable=\y,red]  plot ({\y},{\y});
\draw[red] (2.5,1) node[right]{$At-B$};
\draw[red] (1.2,1.2) node[right]{$At$};
\end{tikzpicture}
\end{center}
\caption{Bounding $S_{\nu,k}$ from above and below by tangent line at infinity}
\label{fig:fig3}
\end{figure}

Besides the upper bound \eqref{upperb} on the partition function, we can construct the lower bound. At the minimum the value of the functional $S_{\nu,k}$ is zero. Therefore its graph is bounded from above by the straight line with the slope $N_ka_{N_k}$, passing through $t=0$, as shown in Fig.\ \eqref{fig:fig3}. 

This immediately translates into the following lower and upper bounds on the partition function
\begin{eqnarray}
\label{ulb}\nonumber
&&N_k!\int_{\mu_1\leqslant\ldots\leqslant\mu_{N_k}} e^{-\gamma_kN_k\mu_{N_k}} \delta\left({\sum}_i\mu_i\right) \Delta^2(e^\mu)\prod_{i=1}^{N_k}d\mu_i\leqslant\\&&\leqslant Z_k(\nu,\omega_0;\gamma_k)\leqslant\\\nonumber
&&\leqslant 
N_k!e^{kN_k\gamma_kC}\int_{\mu_1\leqslant\ldots\leqslant\mu_{N_k}} e^{-\gamma_kN_k\mu_{N_k}} \delta\left({\sum}_i\mu_i\right) \Delta^2(e^\mu)\prod_{i=1}^{N_k}d\mu_i
\end{eqnarray}
The Selberg-type integral here can be done explicitly
\begin{equation}
\int_{\mu_1\leqslant\ldots\leqslant\mu_{N_k}} e^{-\gamma_kN_k\mu_{N_k}} \delta\left({\sum}_i\mu_i\right) \Delta^2(e^\mu)\prod_{i=1}^{N_k}d\mu_i=\gamma_k^{1-N_k}\prod_{n=1}^{N_k-1}\left(\frac{\gamma_k^2}{n^2}-1\right)^{n-N_k}.
\end{equation}
Therefore the bound \eqref{ulb} becomes 
\begin{eqnarray}
\label{bound1}
N_k!\,\gamma_k^{1-N_k}\prod_{n=1}^{N_k-1}\left(\frac{\gamma_k^2}{n^2}-1\right)^{n-N_k}&\leqslant& Z_k(\nu,\omega_0;\gamma_k)\leqslant\\
&\leqslant& N_k!\,e^{kN_k\gamma_kC}\gamma_k^{1-N_k}\prod_{n=1}^{N_k-1}\left(\frac{\gamma_k^2}{n^2}-1\right)^{n-N_k}\nonumber
\end{eqnarray}
The critical value of the coupling constant corresponds to the first simple pole at $n=N_k-1$ on both sides of the bound. In particular, this confirms that the value obtained in Eq.\ \eqref{pole} is sharp. 

The analogous bound on $Z^{\epsilon_k}_k$ can be written as
\begin{eqnarray}
\label{bound2}
N_k!\,\left(\frac{\gamma_k}{\epsilon_k}\right)^{1-N_k}\prod_{n=1}^{N_k-1}\left(\frac{\gamma_k^2}{\epsilon_k^2n^2}-1\right)^{n-N_k}&\leqslant& Z^{\epsilon_k}_k(\nu,\omega_0;\gamma_k)\leqslant\\
&\leqslant& N_k!\,e^{kN_k\gamma_kC}\left(\frac{\gamma_k}{\epsilon_k}\right)^{1-N_k}\prod_{n=1}^{N_k-1}\left(\frac{\gamma_k^2}{\epsilon_k^2n^2}-1\right)^{n-N_k}\nonumber
\end{eqnarray}

\subsection{Lower bound with Jensen inequality}

Applying the Jensen inequality to the log-term of the $\nu$-balancing energy
\begin{equation}
\frac1V\int_M\log\frac{\bs_iP_{ij}s_j}{\sum_l|s_l|^2}\;\nu^n\leqslant\log\frac1V\int\frac{\bs_iP_{ij}s_j}{\sum_l|s_l|^2}\;\nu^n=\log \frac{\tr P}{N_k}
\end{equation}
we immediately get another lower bound for the partition function
\begin{equation}
\label{LBJ}
\int_{\bcal_k}e^{-\gamma_kN_k\log\frac{\tr P}{N_k}}\mathcal D_k\phi=e^{\gamma_kN_k\log N_k}\int_{\bcal_k}(\tr P)^{-\gamma_kN_k}\mathcal D_k\phi \leqslant Z_k(\omega_0,\nu;\gamma_k).
\end{equation}
The last integral can be computed explicitly. Applying the Mellin transform 
\begin{equation}
(\tr P)^{-\gamma_kN_k}=\frac1{\Gamma(\gamma_kN_k)}\int_0^\infty x^{\gamma_kN_k-1}e^{-x\,\tr P}dx
\end{equation}
the matrix integral becomes the partition function for the Wishart ensemble, which we read off from \cite{FKZ4}, Eq. (81),
\begin{equation}
\int_{\bcal_k}e^{-x\,\tr P}\delta(\log\det P)[dP]=x^{-N_k^2}\left[\prod_{j=1}^{N_k}\Gamma(j+1)\right]\cdot G_{0,N_k}^{N_k,0}(x^{N_k}|1,2,\ldots,N_k).
\end{equation}
Then for the integral in \eqref{LBJ} we get
\begin{eqnarray}
\nonumber
\int_{\bcal_k}(\tr P)^{-\gamma_kN_k}\mathcal D_k\phi&=&
\frac1{\Gamma(\gamma_kN_k)}\left[\prod_{j=1}^{N_k}\Gamma(j+1)\right]\int_0^\infty dx\, x^{\gamma_kN_k-N_k^2-1}\cdot G_{0,N_k}^{N_k,0}(x^{N_k}|1,2,\ldots,N_k)\\\nonumber&=&
\frac1{N_k\Gamma(\gamma_kN_k)}\left[\prod_{j=1}^{N_k}\Gamma(j+1)\right]\int_0^\infty dx\, y^{\gamma_k-N_k-1}\cdot G_{0,N_k}^{N_k,0}(y|1,2,\ldots,N_k)\\&=&
\frac1{N_k\Gamma(\gamma_kN_k)}\prod_{j=1}^{N_k}\Gamma(j+1)\Gamma(j+\gamma_k-N_k).
\end{eqnarray}
Hence we have another explicit lower bound
\begin{equation}
\label{Jens}
\frac{e^{\gamma_kN_k\log N_k}}{N_k\Gamma(\gamma_kN_k)}\prod_{j=1}^{N_k}\Gamma(j+1)\Gamma(j+\gamma_k-N_k)\leqslant Z_k(\omega_0,\nu;\gamma_k).
\end{equation}
Again, the first simple pole on the left hand side appears at the critical value \eqref{pole}.

\subsection{Large $k$ asymptotics}

Now we study large $k$ of the bound \eqref{bound2}. These depend on the scaling assumptions on $\gamma_k$ and $\epsilon_k$. We study the following two basic choices 
\begin{eqnarray}
\label{case1}
&(1)&\quad\frac{\gamma_k}{\epsilon_k}=N_k^{\alpha},\quad \alpha>1\\\label{case2}
&(2)&\quad\frac{\gamma_k}{\epsilon_k}=N_k-1+\beta,\quad\beta>0,
\end{eqnarray}
where $\alpha$ and $\beta$ are assumed to be of order one. 
In the case $(1)$ the product in \eqref{bound2} tends to
\begin{equation} 
\quad \log\prod_{n=1}^{N_k-1}\left(\frac{\gamma_k^2}{\epsilon_k^2n^2}-1\right)^{n-N_k}=-(\alpha-1)N_k^2\log N_k-\frac32N_k^2+\mathcal O(N_k\log N_k)
\end{equation}
Taking into account that 
$$
\log N_k!\sim \mathcal O(N_k\log N_k),\quad\log \left(\frac{\gamma_k}{\epsilon_k}\right)^{1-N_k}\sim \mathcal O(N_k\log N_k),
$$
we get
\begin{eqnarray}\nonumber
(1)\quad &&-(\alpha-1)N_k^2\log N_k-\frac32N_k^2+\mathcal O(N_k\log N_k)\leqslant\log Z^{\epsilon_k}_k(\nu,\omega_0;\gamma_k) \leqslant\\
&&\leqslant CkN_k\gamma_k-(\alpha-1)N_k^2\log N_k-\frac32N_k^2+\mathcal O(N_k\log N_k)
\end{eqnarray}
Recall that $C$ is strictly positive constant of order one. To proceed, we shall specify the scaling of $\gamma_k$. First,
\begin{equation} 
(1)'.\;\mbox{if} \;\gamma_k<\frac{(\alpha-1)N_k}{Ck},\quad \mbox{then}\;\log Z^{\epsilon_k}_k(\nu,\omega_0;\gamma_k)\to-c_1N_k^2\log N_k,\quad \mbox{as}\;k\to\infty,
\end{equation}
where $c_1>0$ is a constant i.e. the partition function tends to zero exponentially with the speed as above. The second option
\begin{equation}
(1)''.\;\mbox{if} \;\gamma_k>\frac{(\alpha-1)N_k}{Ck},\quad \mbox{then}\:
\log Z^{\epsilon_k}_k(\nu,\omega_0;\gamma_k)\leqslant CkN_k\gamma_k, \quad\mbox{as}\;k\to\infty,
\end{equation}
i.e. in this case we can only determine the upper bound on large $k$ behavior.

Now we turn to the case (2) in Eq. \eqref{case2}. In this case we have
\begin{equation} 
\quad \log\prod_{n=1}^{N_k-1}\left(\frac{\gamma_k^2}{\epsilon_k^2n^2}-1\right)^{n-N_k}=-N_k^2\log4+\mathcal O(N_k\log N_k)
\end{equation}
and we get the following asymptotic bounds
\begin{eqnarray}
\label{log4}
\nonumber
(2)\quad &&-N_k^2\log4+\mathcal O(N_k\log N_k)\leqslant\log Z^{\epsilon_k}_k(\nu,\omega_0;\gamma_k) \leqslant\\
&&\leqslant CkN_k\gamma_k-N_k^2\log4+\mathcal O(N_k\log N_k)
\end{eqnarray}
We again have two options: first,
\begin{equation} 
(2)'.\;\mbox{if} \;\gamma_k<\frac{N_k}{Ck}\log4,\quad \mbox{then}\;\log Z^{\epsilon_k}_k(\nu,\omega_0;\gamma_k)\to-c_2N_k^2, \quad\mbox{as}\;k\to\infty,
\end{equation}
where, $c_2>0$, and the partition function again tends to zero exponentially. Second option,
\begin{equation} 
(2)''.\;\mbox{if} \;\gamma_k>\frac{N_k}{Ck}\log4,\quad \mbox{then}\;\log Z^{\epsilon_k}_k(\nu,\omega_0;\gamma_k)\leqslant CkN_k\gamma_k, \quad\mbox{as}\;k\to\infty,
\end{equation}
gives the upper bound on large $k$ behavior.

A slightly better asymptotic lower bound follows from the Jensen bound \eqref{Jens}. For the scaling \eqref{case2} we have the following asymptotics
\begin{equation}
\log\frac{e^{\gamma_kN_k\log N_k}}{N_k\Gamma(\gamma_kN_k)}\prod_{j=1}^{N_k}\Gamma(j+1)\Gamma(j+\gamma_k-N_k)=
-\frac12N_k^2+\mathcal O(N_k\log N_k).
\end{equation}
Therefore the constant $-\log 4$ in the bound \eqref{log4} can be replaced by $-1/2$.

\section{The probability measures $d\mu_k$ and the growth of $Z_k$}

Once we have determined $\gamma_k^{\epsilon_k, \rm crit}$ we want to study the sequence of finite measures
\begin{equation}
\label{IPIb} \hat{\mu}_k : = 
e^{-\gamma_kS_{\nu,k}(\omega_0,\phi(P))}\;\mathcal D_k^{\epsilon_k}\phi
\end{equation}
on $\bcal_k$ for a sequence of $\gamma_k^{\epsilon_k} \geq \gamma_k^{\epsilon_k, \rm crit}. $  We also normalize them to obtain a sequence
of probability measures
\begin{equation} \label{muk} \mu_k = \frac{1}{Z^{\epsilon_k}_k(\nu, \omega_0; \gamma_k)}  \hat{\mu}_k. \end{equation}
Since $\bcal_k \subset \kcalomega$, the $\mu_k$ may be regarded as a sequence of probability measures on $\kcalomega$.
Our ultimate goal is to determine the  large $k$ asymptotics of the sequence  and
its dependence  on 
the sequence of coupling constants $\{\gamma_k^{\epsilon_k}\}$ and scaling parameters $\{\epsilon_k\}$. The sequence $\gamma_k^{\epsilon_k, \rm crit}$ probes $\bcal_k$
further out than for larger $\gamma_k$. The definition of `further' depends on the normalization of the symmetric space metrics, i.e.
on the $\epsilon_k$. 
As discussed in \S \ref{SCALINGSECT}, one natural choice is to rescale by
$\epsilon_k = k^{-1} N_k^{-1/2} \simeq k^{-1-n/2}$ (when $\dim M = n$), since this
sequence of symmetric space  tends to the limit Mabuchi metric on $\kcalomega$, see Theorem 1.1 of \cite{CS}.

It is very difficult to determine the asymptotics of the integrals \eqref{MPI}. In this section,
we have a more modest goal: we would like to determine sequences of $\epsilon_k$
 which  `balance' the action term $S_k$ and
the logarithm of the volume form as $k \to \infty$, i.e. so that their contributions to the
Boltzmann weight is of the same order of magnitude. This illuminates the relative contributions
of the action and of the Haar volume form, which is often viewed as the balancing of `energy'
and `entropy'.   We are not able to do this for the integrals \eqref{MPI}  but we are able to obtain
strong results for a simpler integral whose action is a natural and common lower bound for the stability function.

Namely,  we return to \eqref{asymp} and observe that if we rescale $\mu_i\to\mu_i/\epsilon_k$, then for all $(t, \mu, U)$ we have, 
\begin{equation}
\label{asymp2}
S_{\nu,k}(\omega_0,\phi(U^\dagger e^{\mu/\epsilon_k}U)) \geq  \epsilon_k^{-1} N_ka_{N_k}t+\frac{N_k}V\int_M\log \frac{\sum_{i=N_k-r}^{N_k}(\bs U^\dagger)_i (Us)_i}{\sum_l|s_l|^2}\; \nu^n,
\end{equation}
and recall that $r\leqslant N_k-1$ denotes the multiplicity of the  highest eigenvalue $a_{N_k}$. We note that
the $\epsilon_k^{-1}$ appears only in front of the $a_{N_k}$ and not in front of the logarithm of the sum of squares of sections. Since almost
all $a_i$ have multiplicity one, we may regard the right side of \eqref{asymp2} as defining a functional
which depends only on the eigensection $(Us)_{N_k}$ with the highest eigenvalue and we denote it by
\begin{equation} \label{S0} S^0_{\nu, k}(\omega_0, (Us)_{N_k},  \mu/\epsilon_k) :  =   \epsilon_k^{-1} N_k \max_j \{\mu_j\}+\frac{N_k}V\int_M\log \frac{(\bs U^\dagger)_{N_k} (Us)_{N_k}}{\sum_l|s_l|^2}\; \nu^n.
\end{equation}
Then,  for any positive function $\psi$ on $\bcal_k,$
\begin{eqnarray}\label{S0int} \nonumber 
&&\int_{\bcal_k} \psi e^{-\gamma_kS_{\nu,k}(\omega_0,\phi(P))}\mathcal D^{\epsilon_k}\phi 
\leq  \int_{\bcal_k} \psi \; e^{-\gamma_kS^0_{\nu, k}(\omega_0, (Us)_{N_k},  \mu) }\mathcal D_k^{\epsilon_k}\phi \\
&&=  \int_{SU(N_k) \times \R^{N_k}} \psi \; e^{-\gamma_kS^0_{\nu, k}(\omega_0, (Us)_{N_k},\, \mu/\epsilon_k)}\, [d U]\, \delta\left({\sum}_i\mu_i\right)\Delta^2(e^{\mu})[d\mu].  
 \end{eqnarray}
 We refer to \S \ref{SCALINGSECT} for the notation.

For the remainder of this section we consider the simplified integral on the right side of \eqref{S0int}. Our goal is
to determine the $\epsilon_k, \gamma_k$ for which this integral satisfies a large deviations principle. 
This is a mathematically precise way of stating the heuristic idea that the integral tends to an infinite dimensional path integral. In future work, we hope to obtain LDP's for the original measures
\eqref{IPIb}, but we expect this to be a very difficult problem and  view it as a long term
goal.

The integration over $SU(N_k)$ in \eqref{S0int} depends only the first column of $U$ and therefore may be replaced
by an integral over the space $S H^0(M, L^k)$ of  unit norm  holomorphic sections. In fact the functional $S^0_{\nu, k}$
is invariant under scaling  a section by $e^{i \theta}$ and thus the integral descends to the projective space
$\PP H^0(M, L^k)$. It follows that the right side of \eqref{S0int} equals
\begin{equation} \label{S0int2} 
  \int_{\PP H^0(M, L^k) \times \R^{N_k}} \tilde{\psi}(\mu, s) e^{-\gamma_k \left(\epsilon_k^{-1} N_k  \max_j\{\mu_j\} + N_k\frac1V \int_M \log \frac{ ||s(z)||^2_{h^k}}{{\sum_l|s_l|^2}}\nu \right)}  \dcal_{FS_k} (s) \delta\left({\sum}_i\mu_i\right)\Delta^2(e^{\mu})[d\mu]. 
 \end{equation}
Here, $\tilde{\psi}$ is the average of $\psi$ under the map sending $U$ to its first column, and
$ \dcal_{FS_k} $ is the volume form of the Fubini-Study metric on $\PP H^0(M, L^k)$. 

We observe that the integral splits into a product of one over $\R^{N_k}$ and one over $\PP H^0(M, L^k) $. 
We then rewrite both integrals in terms of empirical measures as in \cite{ZZ}. This is a standard technique
in the theory of large deviations. In effect it replaces integration over the changing spaces $\R^{N_k}$ and
$\CP^{N_k-1}$ by integration over a fixed `limit space'. In the case of $\R^{N_k}$, the limit space is the  convex set of $\mcal_0(\R)$ of probability
measures $m$ on $\R$ for which $\int_{\R} x dm(x) = 0$. In the case of $\CP^{N_k}$, it is the space $\mcal_1(M)$  probability
measures on $M$.  One might have expected the limit space to be $\kcalomega$ and clearly there is a relation
with $\mcal_0(\R) \times \mcal_1(M)$. The different limit space is presumably due to the fact
that we have replaced the stability function by the lower bound \eqref{asymp2}.

The integral over $\mu$ may be rewritten in terms of the empirical measure
$$d m_{\mu} : = \frac{1}{N_k} \sum_{j = 1}^{N_k} \delta_{ \mu_j}.$$
Then $ \max_j \{\mu_j\} = ||d m_{\mu}||_{\rm supp}.$
Here, supp $m$ is the support of the measure $m$, i.e. the closure of the set where $m \not= 0$. We also denote by
$ ||d m_{\mu}||_{\rm supp}$ the maximum of $x \in {\rm supp}\, m_\mu$,
$$ ||d m_{\mu}||_{\rm supp} = \max \mbox{supp} \;m_{\mu} = \inf\{x > 0: m_{\mu}[(x, \infty) ]= 0\}. $$
It follows that
$$\Delta^2(e^{\mu}) = e^{2\log \Delta (e^{\mu})} = e^{ N_k^2 \int_{\R \times \R \backslash \rm diag}  \log |e^x - e^y| dm_{\mu}(x)
dm_{\mu}(y)}=e^{N_k^2\Sigma(dm_\mu)}, $$
where $\Sigma$ is the entropy and diag denotes the diagonal.
We can rewrite the integral over $\mu$ in \eqref{S0int2}
as integral over probability measures $m_{\mu}$ on $\R$ with $\int x dm_{\mu} = 0$, with the effective action (rate function),
\begin{equation} \label{RATE}- \gamma_k N_k \epsilon_k^{-1} ||dm_{\mu} ||_{supp} + N_k^2 \int_{\R \times \R}  \log |e^x - e^y| dm_{\mu}(x)
dm_{\mu}(y).  \end{equation}
The two terms here balance as long
as 
\begin{equation} \label{BALANCE1} \gamma_k N_k \epsilon_k^{-1} \approx N_k^2, \end{equation}
where $\approx$ means of the same order of magnitude in terms of powers of $k$.
This happens  exactly for the scaling, considered in \eqref{case2}, when $\gamma_k/\epsilon_k \approx N_k-1+\beta$. In the terminology
of large deviations, the measures \eqref{S0int2}  satisfy a large deviations principal
of speed $N_k^2$ and with the rate function 
\begin{equation} \label{RATE}  I(dm_{\mu}): =   - ||dm_{\mu} ||_{supp} +  \Sigma(d m_{\mu}). \end{equation}
Roughly speaking the logarithmic term is the common  entropy term pushing eigenvalues apart,  while the first term penalizes
measures whose support spreads out too far.

The $U(N_k)$ integral is substantially more complicated, and we only have complete results for Riemann surfaces.
For any $M$,  the integral only involve one column of $U$ and so it descends to an integral over
the projective space of sections,  $\PP H^0(M, L^k)$.
The asymptotics of the logarithm of the $\PP H^0(M, L^k)$ factor of  \eqref{S0int2} were calculated (with different motivations) in \cite{ZZ} in
the case of Riemann surfaces. The main idea is to rewrite the integral over    $\PP H^0(\CP^1, L^k)$ as an integral over the 
$k$th configuration space $(\CP^1)^{(k)}$ consisting  of unordered $k$-tuples of points
$\zeta : = \{\zeta_1, \dots, \zeta_k\}$ on $\CP^1$. The natural map from  $\PP H^0(\CP^1, L^k) \to (\CP^1)^{(k)}$ takes a holomorphic
section to its zero set. It is inverted by sending $\zeta \to  s_{\zeta}$ where
$s_{\zeta} (z) = \prod_{j = 1}^k (z - \zeta_j).$ The change of variables from coefficients relative to an orthonormal basis
to the zero sets puts in the Vandermonde determinant $\Delta(\zeta)$. The Jacobian is calculated in \cite{ZZ}. 
The result is, 
\begin{equation} 
\label{rate}  \int_{(\CP^1)^{k}} 
 e^{-\gamma_k\frac{N_k}V\int_M  \log \frac{ \left| \prod_{j = 1}^k (z - \zeta_j) \right|^2}{\Pi_k(z,z) \qcal(\zeta)} \nu}  \frac{ |\Delta(\zeta)|^2}{\left(\int_{\CP^1} e^{
\int_{\CP^1} G_{h}(z,w) d\mu_{\zeta}(w)} d\nu(z) \right)^{k+1}}  \prod_{j=1}^k d \zeta_j \wedge d \bar{\zeta}_j
\end{equation} 

The calculation of the rate function in the case of $\CP^1$ is given  in \cite{ZZ} and we only quote the result here. Define $dm_\zeta=\frac1k\sum_{j=1}^{k}\delta_{\zeta_j}$. The first term, coming from the action in the exponent in \eqref{rate} contributes
\begin{equation}
\label{actionterm}
- \gamma_k N_k k 
\int_{\mathbb{CP}^1} \int_{\mathbb{CP}^1} G(z, w) dm_{\zeta}(w) d\nu(z),
\end{equation}
where $ G(z, w) \sim2\log|z-w|$ is the Green function, whereas the entropy term from the Vandermonde determinant in \eqref{rate} gives
\begin{equation}
\label{entropyterm}
k^2\int_{\mathbb{CP}^1} \int_{\mathbb{CP}^1}  G(z,w) dm_{\zeta}(w) dm_{\zeta}(z). 
\end{equation}
Thus, taking into account the condition \eqref{BALANCE1}, one finds that if 
\begin{equation}
\gamma_k\approx1,\quad\epsilon_k \approx k^{-1},
\end{equation}
then  (in complex dimension one)  the action contribution \eqref{actionterm} balances the entropy term \eqref{entropyterm}, and the large deviation principle holds for the unitary integration part as well. We note that this sequence of $\epsilon_k$ is different from the ones which normalize
Bergman metrics to converge to the Mabuchi metric (as discussed above).

\section{Discussion}

In this paper we study a model of random Bergman metrics with the $\nu$-balanced energy as the action functional. The main result is the calculation of the critical value of the coupling constant \eqref{pole},\eqref{poleep}, with the path integral converging when $\gamma_k>\gamma_k^{\rm crit}$ and diverging $\gamma_k\leqslant\gamma_k^{\rm crit}$. We also study the large $k$ asymptotics of the partition function, and the form of the limiting path integral measure at large $k$ as given by the large deviation principle. In future work we plan to investigate what kind of random metrics and random surfaces emerge in this model in the $k\to\infty$ limit. 
 
The model of random metrics that we consider here belongs to a more general class of models, involving the stability functions on $\bcal_k$ as action functionals. It would be interesting to generalize our analysis to the case of "balancing energy" - the Aubin-Yau action, restricted to $\bcal_k$ \eqref{AY1}. In this case, the analysis will be much more subtle, and interesting, because asymptotics of the action at geodesic infinity are more complicated \cite{PST,PS3} and also because it is tricky to get a handle on the asymptotic behavior of $y$-intercept. On the other hand, let us also mention that the lower bound measures \eqref{ulb} and \eqref{LBJ}, by construction are well-defined eigenvalue-type measures on $\bcal_k$, exactly solvable in the matrix-model sense. It would be interesting to study what kind of random metrics emerge from these measures.

A different, and possibly unifying point of view on the geometric actions can be formulated as follows. Consider a representation $\rho$ of $G=SL(N_k,\mathbb C)$ on a very big ($\dim V\gg N_k$) vector space $V$. Take a vector $v\in V$, called Chow vector, and $A\in G$. Interesting geometric functionals, restricted to $\bcal_k$ (recall, that $P=A^\dagger A\in \bcal_k$) all appear to have the following abstract, yet completely transparent form
\begin{equation}
S_{\rm geom}(A)=\log ||\rho(A)\cdot v||,
\end{equation}
or a combination of logarithms of this kind, see \cite{T0} and \cite{PST,PS1,PS2,PS}. The path integral measure $\mu_k$ with the action as above, with the help of the Mellin transform can be written as
\begin{equation}
\int_{\bcal_k}\mu_k=\int_{\bcal_k}e^{-\gamma_k\log ||\rho(A)\cdot v||}\mathcal D_k\phi=\frac1{\Gamma(\gamma_k)}\int_0^{\infty} x^{\gamma_k-1}dx\int_{\bcal_k}e^{-x(v^*\rho(A)^*\rho(A)\cdot v)}\mathcal D_k\phi.
\end{equation}
The unitary part of the last matrix integral can be thought of as a generalized Itzykson-Zuber integral, where instead of a pair of  hermitian matrices, one considers a big (of the size $\gg N_k$) hermitian matrix $\rho(A^\dagger)\rho(A)$ of the representation $\rho$ and a rank one hermitian matrix $H_{ij}=v_iv^*_j$ of the same size. It would be interesting to study this type of matrix integrals.

We plan to address this set of questions in future work. 
\vspace{.3cm}

{\bf Acknowledgments}. We thank J.~Sturm for many discussions of stability functions and stability integrals, J.~Fine for discussions of convergence of metrics, and D.~H.~Phong for encouragement to study integration over Bergman metrics. SK is supported by the postdoctoral fellowship from the Alexander von Humboldt Foundation. He is also supported in part by the DFG-grant ZI 513/2-1, grants RFBR 14-01-00547, NSh-3349.2012.2, and by Ministry of Education and Science of the Russian Federation under the contract 8207. SZ is partially supported by NSF grant DMS-1206527.

\end{document}